\documentclass{article}
\usepackage{spconf,amsmath,graphicx,siunitx,microtype,amsfonts}
\usepackage[numbers]{natbib}

\usepackage{amsmath,amsfonts,bm}

\def\eqref#1{equation~\ref{#1}}

\def\1{\bm{1}}

\def\vr{{\bm{r}}}

\def\vy{{\bm{y}}}

\DeclareMathAlphabet{\mathsfit}{\encodingdefault}{\sfdefault}{m}{sl}
\SetMathAlphabet{\mathsfit}{bold}{\encodingdefault}{\sfdefault}{bx}{n}

\usepackage{array}
\usepackage{booktabs}

\usepackage{microtype}
\usepackage{subfig}
\usepackage{graphicx,wrapfig}
\usepackage{url}
\usepackage{amsmath}
\usepackage{amsfonts,amssymb}
\usepackage{siunitx}

\usepackage{color}
\usepackage{multirow}

\usepackage{booktabs}
\usepackage[table,xcdraw]{xcolor}
\usepackage[colorlinks,linkcolor=black,urlcolor=black,citecolor=black]{hyperref}

\usepackage{xcolor}
\usepackage{algpseudocode}
\usepackage{siunitx}
\usepackage{multirow}
\usepackage{amsmath}
\usepackage{csquotes}
\usepackage{tabularx} 
\usepackage[linesnumbered,ruled,noend,noline]{algorithm2e}

\newcommand{\vSampleNumber}{N}
\newcommand{\vSampleNumberIndex}{n}
\newcommand{\vClassNum}{C}
\newcommand{\vClassNumindex}{c}
\newcommand{\vDataAudio}{\mathbf{X}}
\newcommand{\vDataLabel}{\mathbf{Y}}
\newcommand{\vWeightOmAP}{\mathbf{W}}

\newcommand{\vEvalMetricValue}{z}
\newcommand{\vDataLabelPrediction}{\hat{\mathbf{Y}}}
\newcommand{\vClassifier}{F}
\newcommand{\vEvaluaionMetric}{\operatorname{Eval}}
\newcommand{\vDataset}{\mathbb{D}}
\newcommand{\vGraphNodeDistance}{\mathbf{D}}

\newcommand{\vTPS}{\operatorname{TP}}
\newcommand{\vFNS}{\operatorname{FN}}
\newcommand{\vFPS}{\operatorname{FP}}
\newcommand{\vequalsign}{~\texttt{=}~}
\newcommand{\vequalsignnospace}{\texttt{=}}
\newcommand{\vRecallMatrix}{\mathbf{R}}
\newcommand{\vPrecisionMatrix}{\mathbf{P}}

\newcommand{\vGraph}{\mathcal{G}}
\newcommand{\vEdge}{E}
\newcommand{\vNode}{V}

\ninept

\title{Ontology-aware learning and evaluation for audio tagging}
\name{
      Haohe Liu$^{1}$,
      Qiuqiang Kong$^{2}$,
      Xubo Liu$^{1}$,
      Xinhao Mei$^{1}$,
      Wenwu Wang$^{1}$,
      Mark D. Plumbley$^{1}$
      }
\address{
    $^1$Centre for Vision, Speech and Signal Processing~(CVSSP), University of Surrey, UK \\
    $^2$Speech, Audio, and Music Intelligence~(SAMI) Group, ByteDance, China\\
}

\begin{document}
%

\maketitle

\begin{abstract}
This study defines a new evaluation metric for audio tagging tasks to overcome the limitation of the conventional mean average precision~(mAP) metric, which treats different kinds of sound as independent classes without considering their relations.
Also, due to the ambiguities in sound labeling, the labels in the training and evaluation set are not guaranteed to be accurate and exhaustive, which poses challenges for robust evaluation with mAP. The proposed metric, ontology-aware mean average precision~(OmAP) addresses the weaknesses of mAP by utilizing the AudioSet ontology information during the evaluation. Specifically, we reweight the false positive events in the model prediction based on the ontology graph distance to the target classes. The OmAP measure also provides more insights into model performance by evaluations with different coarse-grained levels in the ontology graph. We conduct human evaluations and demonstrate that OmAP is more consistent with human perception than mAP. To further verify the importance of utilizing the ontology information, we also propose a novel loss function~(OBCE) that reweights binary cross entropy~(BCE) loss based on the ontology distance. Our experiment shows that OBCE can improve both mAP and OmAP metrics on the AudioSet tagging task. Our code is open-sourced\footnote{\url{github.com/haoheliu/ontology-aware-audio-tagging}}.
\end{abstract}

\begin{keywords}
machine learning, audio tagging, ontology, evaluation metric
\end{keywords}

\vspace{-4mm}
\section{Introduction}
\label{sec:intro}
Audio tagging is a task that tags an audio clip with one or more labels. Audio tagging has attracted increasing interest from researchers in recent years~\cite{kong2020panns, gong2021psla}, with the increasing number of papers in the Detection and Classification of Acoustic Scenes and Events (DCASE) data challenges~\cite{giannoulis2013detection, mesaros2017detection, mesaros2017dcase,liu2022segment}. Audio tagging has several applications such as urban noise control~\cite{bello2019sonyc}, audio retrieval~\cite{guo2003content}, and audio monitoring~\cite{ward_dawes_2021}. 

Most evaluation metrics for audio tagging systems are based on the confusion matrix~\cite{branco2016survey}. Early works~\cite{cakir2016domestic, kong2017joint} employ the metrics such as the equal error rate~(EER), and sensitivity index~\cite{wickens2001elementary}. There are also metrics like F-score~\cite{informationretrieval1979}. However, the need for choosing a suitable threshold for F-score makes it less straightforward to use. Many recent studies adopt the mean average precision~(mAP) as the evaluation metric for audio tagging~\cite{gong2021ast,kong2020panns,liu2022simple}, which measures the area under the precision-recall curve. Based on a simple two-level hierarchical class ontology, Bello et al.~\cite{bello2019sonyc} proposed to calculate both fine-grained and coarse-grained mAP for evaluation to investigate the trade-off between fine-grained, potentially erroneous labels and coarse-grained, likely accurate labels. 
The mAP metric is preferable on datasets with unbalanced class distribution~\cite{davis2006relationship}, such as AudioSet~\cite{gemmeke2017audio}.  

Recently, a number of large scale dataset for audio tagging have been proposed, such as AudioSet~\cite{gemmeke2017audio}, and FSD50K~\cite{fonseca2021fsd50k}. There are \num{527} classes in AudioSet and \num{200} classes in FSD50K, with an unbalanced distribution of total duration in each class. To address the duration imbalance issues~\cite{gong2021psla}, the primary evaluation metric on these datasets is the class-wise mAP, in which the average precision~(AP) score is calculated for each class and averaged as the final result. 
On calculating the mAP, if a predicted sound event does not appear in the target labels, the prediction will be considered as false positive~(FP). Otherwise, it will be counted as a true positive~(TP). 
However, calculating FP in this way has the following problems:


\begin{figure}[tbp] 
  \centering
  \includegraphics[page=1,width=0.99\linewidth]{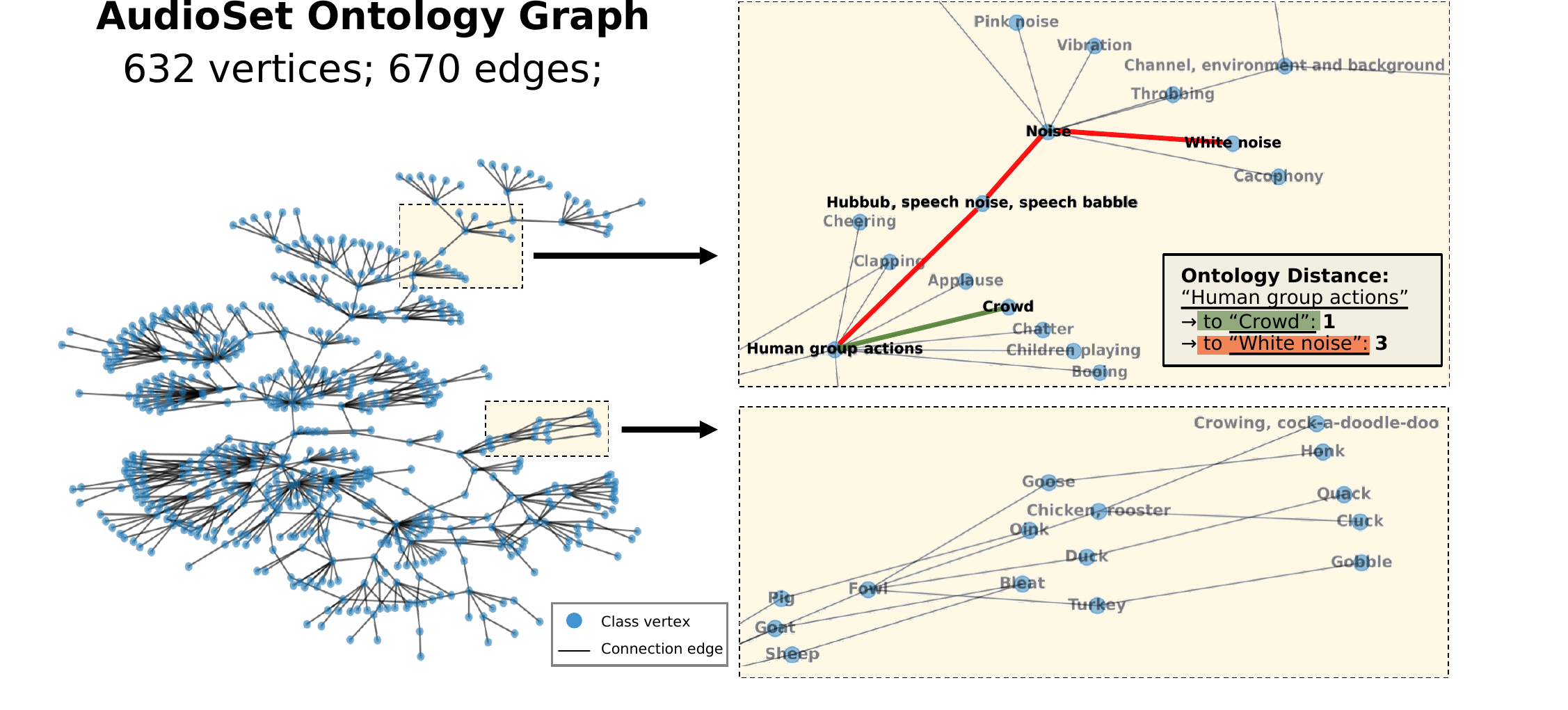}
  \caption{The ontology graph of the AudioSet~\cite{gemmeke2017audio} classes.}
  \label{fig-ontology}
  \vspace{-6mm}
\end{figure}

\textit{Missing labels in the dataset: } The labels in the audio tagging dataset are not always correct and may contain missing labels. For example, it is estimated that more than 50\% of the labels for around $30\%$ of the classes in AudioSet are in-correct\footnote{\url{https://research.google.com/audioset/dataset/index.html}}. In the evaluation set of AudioSet, there are $4895$ files containing the label \textit{Speech}, while only $2.1\%$ of them have label \textit{Male Speech} or \textit{Female Speech}. 
In this case, even if the model learned to estimate gender on all files with speech, $97.9\%$ of the gender labels will be considered to be false positive. In fact, compared to the \textit{Speech} class with an AP of $0.80$, the experiment in~\cite{gong2021psla} shows that the APs for \textit{Male Speech} and \textit{Female Speech} are only $0.07$ and $0.09$, respectively.

\textit{The non-exclusive nature of sound classes: } Sound classes are not always mutually independent. There are also inclusiveness~(e.g., \textit{Music} and \textit{Guitar}) or intersections~(e.g., \textit{Shout} and \textit{Yell}) relations between different sounds. Therefore, we believe FPs should be reweighted by their ``seriousness''. For example, if the target label is the \textit{Giggle} sound, the intuition is that an FP prediction \textit{Laughter} is less ``serious'' than an FP prediction \textit{Guitar}, because \textit{Giggle} is semantically closer to \textit{Laughter}. Previous evaluation methods fail to consider these class-level relations and may not ideally reflect the model performance.

We propose an improved metric: ontology-aware mean average precision~(OmAP) to address the above problems. OmAP reweights the FP predictions based on the ontology distance between the prediction and the target labels~(see Figure~\ref{fig-ontology}).
In this way, the FPs can be reweighted based on the relation between the predicted and the target labels. 
By grouping the classes based on the ontology distance, we also show that OmAP can be calculated at different coarse-grained levels, reflecting a more thorough view of the model performance. We conduct human evaluations to study the consistency of different objective metrics to human perceptions and observe higher consistency of the proposed metric than the conventional mAP.
With similar motivation as OmAP, we also propose a novel ontology-aware binary cross entropy~(OBCE) loss function to train audio tagging systems. OBCE reweights the binary cross entropy loss~(BCE) based on the class ontology. 
Our experiments show that the OBCE loss can not only improve the OmAP but also improve mAP, which further indicates that the ontology information is useful for model optimization. 


\section{Problem formulation}
\label{sec-problem-formulation}

\label{sec-audio-tagging}
\textbf{Audio tagging}~~
Let $\mathbb{\vDataset}_{\text{train}} \vequalsignnospace \langle \vDataAudio^{\prime}_{\vSampleNumber \times T}, \vDataLabel^{\prime}_{\vSampleNumber\times \vClassNum} \rangle$ denotes an audio tagging training dataset, where $\vDataAudio^{\prime}$ and $\vDataLabel^{\prime}$ denote $\vSampleNumber$ audio samples and their labels. 
The audio sample length and the total number of classes are denoted by $T$ and $\vClassNum$, respectively. 
We define class as a particular type of sound, and define label(s) as the class(es) that appeared in an audio sample. Each audio sample can have one or multiple labels.
The label matrix $\vDataLabel^{\prime}$ only has elements with value zero and one. If the element $\vDataLabel^{\prime}_{\vSampleNumberIndex, i}$ equals to one, that indicates the $n$-th sample is labeled with the $i$-th class.
The label of the $n$-th sample is given by $L_{n}\vequalsignnospace \left\{i~|~\vDataLabel^{\prime}_{\vSampleNumberIndex, i}\vequalsign1\right\}$. 
The audio tagging model $\vClassifier(\cdot)$ aims to estimate $\hat{\vDataLabel}^{\prime} \vequalsign \vClassifier(\vDataAudio^{\prime})$, where $\hat{\vDataLabel}^{\prime}\in[0,1]_{\vSampleNumber \times \vClassNum}$ is the estimation of $\vDataLabel^{\prime}$. 
The performance of the audio tagging model $\vClassifier(\cdot)$ is calculated on the evaluation dataset $\mathbb{\vDataset}_{\operatorname{eval}} \vequalsignnospace \langle \vDataAudio, \vDataLabel \rangle$, formulated as $\vEvalMetricValue\vequalsign\vEvaluaionMetric(\vClassifier(\vDataAudio), \vDataLabel)$, where $\vEvaluaionMetric(\cdot)$ is an evaluation metric. 

\label{sec-mean-average-precision}
\noindent
\textbf{Mean average precision}~~
Mean average precision~(mAP) is a metric that has been widely used in audio tagging~\cite{kong2020panns,gong2021psla} and image object detection~\cite{zhao2019object} tasks. Let $z$ denote the mAP value. The calculation of mAP is based on the average AP of each class $c$, given by 

\vspace{-2mm}
\begin{equation}
    \vEvalMetricValue\vequalsign\sum_{\vClassNumindex\vequalsign1}^{\vClassNum}\frac{\vEvalMetricValue_{c}}{\vClassNum},~~\vEvalMetricValue_{c}\vequalsign\mathcal{P}(\vDataLabel_{:,c}, \vDataLabelPrediction_{:,c})\vequalsign\mathcal{A}(\vPrecisionMatrix_{:,c},\vRecallMatrix_{:,c}),
    \label{eq-map}
\end{equation}
\vspace{-2mm}

where $z_{c}$ is the AP for class $c$, $\mathcal{P}(\cdot)$ is the function for calculating AP, and $\mathcal{A}(\cdot)$ denotes the function that calculates the area under curve~\cite{davis2006relationship}. We use $\vPrecisionMatrix$ and $\vRecallMatrix$ to denote the precision and recall matrix, respectively. The shape of $\vPrecisionMatrix$ and $\vRecallMatrix$ is $N\times C$ because we calculate the precision and recall on $N$ different thresholds and $C$ classes. The $N$ thresholds for a class $c$ are the $N$ values in the label estimation $\vDataLabelPrediction_{:,\vClassNumindex}$~\cite{kong2020panns,gong2021psla}.
The AP for class $c$ is calculated by the area under the precision-recall curve formed by $N$ pairs of precision and recall coordinates $(\vPrecisionMatrix_{:,c},\vRecallMatrix_{:,c})\vequalsign(P_{\vSampleNumberIndex, \vClassNumindex}, R_{\vSampleNumberIndex, \vClassNumindex})_{n\vequalsign{1,2,...,N}}$.
Given a threshold $\gamma\vequalsign\vDataLabelPrediction_{n,\vClassNumindex}$, the coordinates are calculated by
\begin{equation}
    (P_{\vSampleNumberIndex, \vClassNumindex}, R_{\vSampleNumberIndex, \vClassNumindex})
    \vequalsign
    (\frac{\vTPS_{\vSampleNumberIndex,\vClassNumindex}}{\vTPS_{\vSampleNumberIndex,\vClassNumindex}+\vFPS_{\vSampleNumberIndex,\vClassNumindex}},
    \frac{\vTPS_{\vSampleNumberIndex,\vClassNumindex}}{\vTPS_{\vSampleNumberIndex,\vClassNumindex}+\vFNS_{\vSampleNumberIndex,\vClassNumindex}})
    \label{eq-precision-recall-map}
\end{equation}
where $\vTPS_{\vSampleNumberIndex,\vClassNumindex}$ and $\vFPS_{\vSampleNumberIndex,\vClassNumindex}$ 
are formulated as $|\{{i~|~\vDataLabelPrediction_{i, \vClassNumindex} \texttt{>} \gamma, \vDataLabel_{i, \vClassNumindex}\vequalsign1}\}|$ and $|\{{i~|~\vDataLabelPrediction_{i, \vClassNumindex} \texttt{>} \gamma, \vDataLabel_{i, \vClassNumindex}\vequalsign0}\}|$, respectively, where $|\cdot|$ denotes the size of a set.
In Equation~(\ref{eq-precision-recall-map}), the denominator of $R_{\vSampleNumberIndex, \vClassNumindex}$, 
$\vTPS_{\vSampleNumberIndex,\vClassNumindex}+\vFNS_{\vSampleNumberIndex,\vClassNumindex}$, is a constant and equal to the total number of positive labels, $\sum_{i\vequalsign1}^{N}\vDataLabel_{i, \vClassNumindex}$, for threshold $\gamma$ and class $c$.


\label{sec-ontology-graph}
\noindent
\textbf{Audio class ontology}~~
The $\vClassNum$ audio classes can be represented by an undirected complete graph $\vGraph\vequalsign( \vNode, \vEdge )$, where $\vNode$ and $\vEdge$ denote sets for vertices and edges, respectively. 
We use $v_c \in \vNode$ to denote the vertex for class $c$. We define the minimum distance between two vertices $v_{i}$ and $v_{j}$, $\vGraphNodeDistance_{i,j}$, as the smallest number of edges to connect $v_{i}$ and $v_{j}$, given by $\vGraphNodeDistance_{i,j}\vequalsign\operatorname{Dist}(v_{i}, v_{j})$, where $\operatorname{Dist}(\cdot)$ is the minimum distance calculation function.
The distance matrix $\vGraphNodeDistance\in \mathbb{Z}^{+}_{C\times C}$ is symmetric with shape $\vClassNum \times \vClassNum$. We also refer to the graph $\vGraph$ as the ontology. One of the most comprehensive audio class ontologies is the one proposed by AudioSet~\cite{gemmeke2017audio}. 

\section{Ontology-aware mean average precision}
\label{sec-method}

\label{sec-new-metric}

As discussed in Section~\ref{sec:intro}, evaluating an audio tagging system with mAP is subject to the missing label problem and will overlook the relations between classes. Our proposed OmAP addresses these problems by incorporating the ontology graph into the evaluation process. Motivated by~\cite{bello2019sonyc}, we design OmAP to evaluate model performance on multiple coarse-grained levels $\lambda$ to gain more insights into the model performance. The final OmAP $z$ is the mean value of ontology-aware average precision~(OAP) on different class $c$ and $\lambda$, defined by
\begin{equation}
    \vEvalMetricValue\vequalsign\sum_{\lambda\vequalsign0}^{D_{\operatorname{m}}}\sum_{\vClassNumindex\vequalsign1}^{\vClassNum}\frac{\vEvalMetricValue_{\lambda,c}}{D_{\operatorname{m}}\vClassNum},~
    \vEvalMetricValue_{\lambda,c}\vequalsign\mathcal{P}^{\prime}(\vDataLabel_{:,c}, \vDataLabelPrediction_{:,c},\lambda,\vGraph)\vequalsign\mathcal{A}(\vPrecisionMatrix_{:,c},\vRecallMatrix_{:,c})
    \label{eq-OmAP}
\end{equation}
where $\mathcal{P^{\prime}(\cdot)}$ denotes the OAP evaluation function, $\lambda$ denotes the coarse-grained level, and $D_{\operatorname{m}}\vequalsignnospace\operatorname{max}(\vGraphNodeDistance)$ is the maximum distance between two arbitrary vertices in $\vGraph$, representing the coarsest level of evaluation. 
We will introduce the detail of multi-level coarse-grained evaluation in Equation~(\ref{eq:dijkstra}).
In a similar way as Equation~(\ref{eq-precision-recall-map}), for each class $c$ with $N$ thresholds
$(\vDataLabelPrediction_{\vSampleNumberIndex,\vClassNumindex})_{n\vequalsign{1,2,...,N}}$, we calculate the $N$ coordinates of the OAP precision-recall curve by
\begin{equation}
    (P_{\vSampleNumberIndex, \vClassNumindex}, R_{\vSampleNumberIndex, \vClassNumindex})
    \vequalsign
    (\frac{\vTPS_{\vSampleNumberIndex,\vClassNumindex}}{\vTPS_{\vSampleNumberIndex,\vClassNumindex}+\vFPS_{\vSampleNumberIndex,\vClassNumindex}\vWeightOmAP_{\vSampleNumberIndex,\vClassNumindex}},
    \frac{\vTPS_{\vSampleNumberIndex,\vClassNumindex}}{\vTPS_{\vSampleNumberIndex,\vClassNumindex}+\vFNS_{\vSampleNumberIndex,\vClassNumindex}}),
    \label{eq-precision-recall-omap}
\end{equation}
in which the calculation of $\vFNS$, $\vFPS$, and $\vTPS$ are the same as Equation~(\ref{eq-precision-recall-map}), and the only difference is the reweight matrix $\vWeightOmAP_{\vSampleNumberIndex,\vClassNumindex}$, which represents how ``serious'' is the mistake if class $c$ appears as an FP on the $n$-th samples. 
The shape of the reweighting matrix $\vWeightOmAP$ is $N\times C$. The value of $\vWeightOmAP_{\vSampleNumberIndex,\vClassNumindex}$ will be small if $\vFPS_{\vSampleNumberIndex,\vClassNumindex}$ represents only a minor mistake.
The seriousness of $\vFPS_{\vSampleNumberIndex,\vClassNumindex}$ is quantified with the ontology graph based on the assumption that a label prediction that is further away from the target label is a more ``serious'' mistake.
To calculate $\vWeightOmAP$, we first quantify the ontology distance $\vGraphNodeDistance$ by


\begin{equation}
    \label{eq:dijkstra}
    \vGraphNodeDistance_{i,j} 
    \begin{cases}
      d_{i,j}, & \text{if}~ d_{i,j} > \lambda \\
      0, & \operatorname{otherwise}
    \end{cases},~~~d_{i,j} \vequalsign \operatorname{Dist}(v_i,v_j)
\end{equation}

\begin{figure}[tbp] 
  \centering
  \vspace{-3mm}
  \includegraphics[page=1,width=0.99\linewidth]{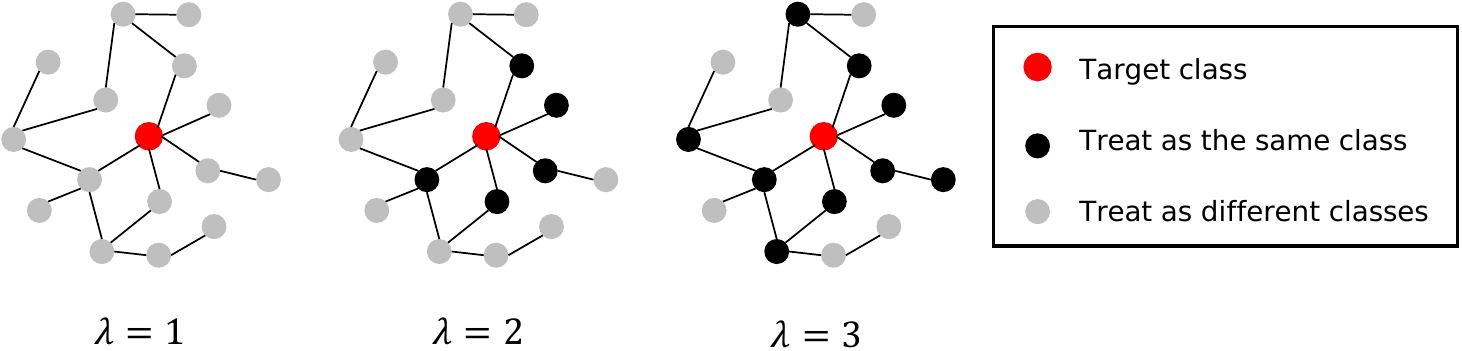}
  \caption{Calculate false positive on different coarse levels $\lambda$ on the node highlighted with red. The black nodes will be treated as the same class as the target class but not false positives.}
  \label{fig-coarse-graph}
  \vspace{-6mm}
\end{figure}

As illustrated in Figure~\ref{fig-coarse-graph}, OmAP is calculated with multiple coarse-grained levels $\lambda$ from $0$ to $D_{m}$, $\lambda \in \mathbb{Z}_{\geq 0}$, where $D_{\operatorname{m}}$ is the maximum distance between two arbitrary vertices in $\vGraph$. 
Evaluation with different $\lambda$ can alleviate the missing label problem because the missed labels, which are more likely to be closer to target labels, can be ignored in certain coarse level.
For example, if $\lambda=2$, the FP on classes that have a minimum distance smaller or equal than two~(e.g.,~\textit{Female Speech}) to the target classes~(e.g.,~\textit{Speech}) will not be taken into account. 
With the distance matrix, we can calculate $\vWeightOmAP_{\vSampleNumberIndex,\vClassNumindex}$ by
\begin{equation}
    \vWeightOmAP_{\vSampleNumberIndex,\vClassNumindex}\vequalsign\frac{1}{\mu}\operatorname{min}\left\{\vGraphNodeDistance_{\vClassNumindex, k}~|~k\in L_{n}\right\},~~\mu~\vequalsignnospace\operatorname{mean}(\vGraphNodeDistance),
    \label{eq-reweight-matrix}
\end{equation}
where $L_{n}$ is the label for the $n$-th sample, function $\operatorname{mean}(\cdot)$ calculates the mean value of all the elements in a matrix, and $\mu$ is the mean value of $\vGraphNodeDistance$. We divide $\vWeightOmAP$ by $\mu$ to ensure 
the value of OmAP can have a similar scale as mAP. The reweighting matrix $\vWeightOmAP$ is dependent on $\vGraphNodeDistance$, which is calculated with different $\lambda$, thus $\vWeightOmAP$ also has different values on different $\lambda$. Finally, $\vWeightOmAP_{\vSampleNumberIndex,\vClassNumindex}$ can be utilized in Equation~(\ref{eq-precision-recall-omap}) and~(\ref{eq-OmAP}) to calculate OmAP at different coarse-grained levels.

\section{Ontology-aware binary cross entropy loss}
\label{sec-loss-function}
We propose an OBCE loss, $L_{\text{obce}}$, to explore if the ontology information is beneficial for model optimization. 
The intuition behind $L_{\text{obce}}$ is similar to OmAP, alleviating the missing-label problem, and treating each class differently according to its distance to the target classes. The proposed OBCE loss is built upon the traditional BCE loss. Given the target and label prediction $\vy$ and $\hat{\vy}$ of an audio sample, the BCE loss can be formulated as
\begin{equation}
    \label{eq-bce-loss}
    \mathcal{L}_{\text{bce}}\vequalsign \operatorname{mean} (\vy\odot\operatorname{log}(\hat{\vy})+(1-\vy)\odot\operatorname{log}(1-\hat{\vy})),
\end{equation}
where $\odot$ means the Hadamard product, $\operatorname{log}$ means element-wise log, and $\vy$ is the label vector with elements of ones and zeros. Compared with $\mathcal{L}_{\text{bce}}$, the OBCE loss reweights the loss function for each class $c$ based on the distance of the predictions to the target labels. 
Based on the similar motivation discussed in Section~\ref{sec-new-metric}, OBCE loss is designed to assign a smaller weight to false predictions that are closer to the target labels. 
Assigning weight to false prediction can also alleviate the missing-label problem. 
As shown in Algorithm~\ref{algorithm:weight}, we calculate the loss weight of class $c$, $\vr_c$, based on the minimum distance between the vertex of class $c$ and vertices of the target label set $L_n$. 
With the loss weight $\vr$, the OBCE loss can be formulated as

\vspace{-4mm}
\begin{algorithm}[tbp]
\footnotesize
    \SetKw{KwBy}{by}    
    \SetKw{KwAnd}{and}    
    \SetKw{KwOr}{or}    
    \SetKwInOut{Input}{Inputs}
    \SetKwInOut{Output}{Output}
    \Input{Ontology $\vGraph$, label for the $n$-th sample $L_n$, total number of classes $C$, distance power factor $\beta$.}
    \Output{Loss weight vector $\vr$ with length $C$.}
    \For{$c ~\operatorname{\mathbf{in}}~ [1,2,...,C]$}{ 
         $\vr_{c}\vequalsign\operatorname{min}\{d^{\beta}~|~d\vequalsignnospace \operatorname{Dist}(v_c,v_k), k\in L_{n} \}$; \\
    }
    $\vr \gets \vr / \operatorname{max}(\vr) ;$  \Comment{\textit{Preparation for line 4.}} \\
    $\vr_{k} \gets 1.0,~k\in L_{n} ;$ \Comment{\textit{Target labels have the highest weight.}}  \\
    \For{$c ~\operatorname{\mathbf{in}}~ [1,2,...,C]$}{  
         $\vr_c \gets \vr_c / \operatorname{mean}(\vr)$;    \\
         \Comment{\textit{Let mean $\bar{\vr}\vequalsignnospace1$. For a fair comparison with the BCE loss.}} \\
    }
    \caption{Calculate OBCE loss weight during training.}\label{algorithm:weight}
\end{algorithm}

\begin{equation}
    \label{eq-obce-loss}
    \mathcal{L}_{\text{obce}}\vequalsign \operatorname{mean} (\vr \odot (\vy\odot\operatorname{log}(\hat{\vy})+(1-\vy)\odot\operatorname{log}(1-\hat{\vy})))
\end{equation}

Note that in OmAP, we calculate the vertices distance in $\vGraph$ by simply calculating the number of edges~(see Equation~(\ref{eq:dijkstra})). 
However, this assumption is not necessarily optimal for model optimization using OBCE. Hence, in Algorithm~\ref{algorithm:weight}, we raise the distance $d$ with the power of a distance power factor $\beta$ to explore the effect of non-linear distance. We empirically observe that $\beta$ can affect the model performance on mAP and OmAP. 
Since the OBCE loss has a higher weight on classes that are further away from the target labels, the OBCE loss tends to emphasize more on coarse-grained classification. Therefore, we use the BCE loss together with the OBCE loss to ensure the model still is sufficiently optimized for fine-grained classification. The final loss function $\mathcal{L}$ is the combination of the BCE and OBCE losses, given by $\mathcal{L}=(\mathcal{L}_{\text{bce}}+\mathcal{L}_{\text{obce}})/2$, in which the division of these two is to ensure a similar scale of $\mathcal{L}$ with $\mathcal{L}_{\text{bce}}$ for fair comparisons in the experiments.

\vspace{-4mm}
\section{Experiments}
\label{sec-experiments}

\begin{table}[tbp]
\centering
\footnotesize
\vspace{-3mm}
\begin{tabular}{cccccc}
\toprule[\heavyrulewidth]
       Model & Params & mAP      & OmAP     & OmAP$_{0}$ \\
\midrule
PANN~\cite{kong2020panns}  & 42~M              & 43.3 & 76.7  & 54.3      \\
PSLA~\cite{gong2021psla}   & 14~M             & 43.7     & 77.6    & 55.3         \\
AST~\cite{gong2021ast}    & 88~M                      & 45.6  & \textbf{78.5} & 57.0      \\
HTS-AT~\cite{chen2022hts} & 31~M          & \textbf{46.4} & \textbf{78.5}  & \textbf{57.7} \\
\midrule
\end{tabular}
\caption{The performance of state-of-the-art methods on AudioSet. We also report the OmAP at the finest-grained level when $\lambda\vequalsignnospace0$, denoted by OmAP$_{0}$. All the metrics are reported in the percentage format.}
\label{tab-sotamethod}
\end{table}

\begin{figure}[tbp] 
  \centering
  \vspace{-4mm}
  \includegraphics[page=1,width=1.0\linewidth]{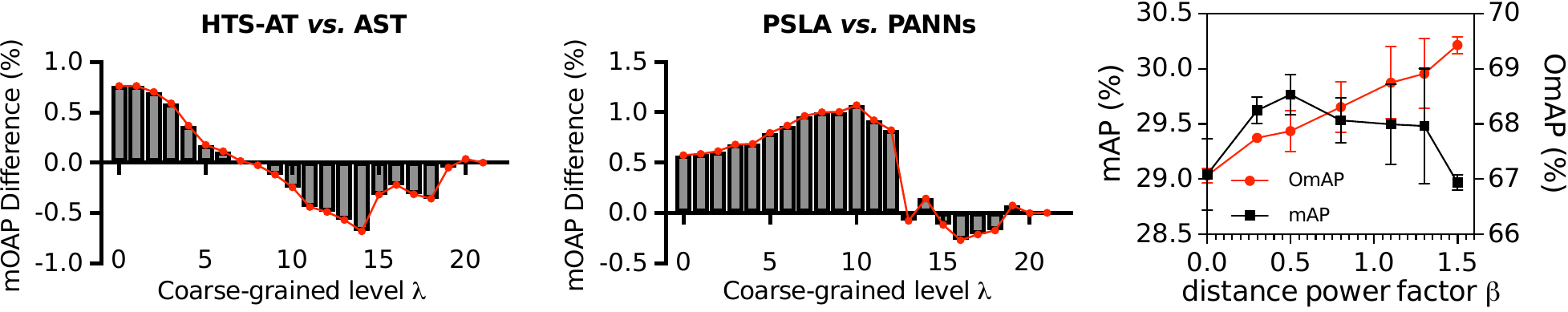}
  \caption{The left two figures show the differences of OmAP at each coarse-grained level between two groups of models. Although HTS-AT and AST have the same OmAP~(see Table~\ref{tab-sotamethod}), HTS-AT is better on smaller coarse-grained levels. The rightmost figure shows the OmAP and mAP performance with different $\beta$ settings in the OBCE loss.}
  \label{fig-sota}
  \vspace{-4mm}
\end{figure}

We conduct experiments on the AudioSet balanced subset~(AudioSet-20K), full AudioSet~(AudioSet-2M)~\cite{gemmeke2017audio}, and the FSD50K dataset~\cite{fonseca2021fsd50k}. All the datasets are resampled into a sampling rate of \num{16} kHz following~\cite{kong2020panns, liu2022learning}. The AudioSet ontology is a complete graph, in which the maximum distance between two nodes is $D_m\vequalsignnospace21$. The FSD50K datasets use part of the AudioSet ontology with \num{200} classes. 
We use the same backbone as~\cite{gong2021psla}, which is an ImageNet pretrained EfficientNet-B2~\cite{tan2019efficientnet} with four-head attentions block. The detailed setup of hyper-parameters is the same as Liu et al.~\cite{liu2022learning}. 
For the distance power factor $\beta$, we use $1.0$ by default, except for the experiments in Figure~\ref{fig-sota}. We also report the OmAP when $\lambda\vequalsign0$, denoted by mAP$_{0}$, which is the most fine-grained evaluation level.

\noindent
\textbf{OmAP of state-of-the-art methods}~~
As shown in Table~\ref{tab-sotamethod}, we evaluate several SOTA methods for audio tagging with both mAP and our proposed OmAP metrics. 
The evaluation is performed on the open-sourced pretrained versions of these four methods.
In Table~\ref{tab-sotamethod} we see that the OmAP score of AST and HTS-AT are the same, while the OmAP of AST when $\lambda\vequalsignnospace0$ is $0.7\%$ lower than HTS-AT, which indicates that HTS-AT and AST do not perform the same on different $\lambda$. 
We further visualize the difference between HTS-AT and AST in Figure~\ref{fig-sota}, which shows that HTS-AT performs better on smaller $\lambda$ while AST performs better on larger $\lambda$. 
This indicates that the hierarchical structure and shifted window attention in HTS-AT~\cite{chen2022hts} might benefit fine-grained classifications. Although all three evaluation metrics show that PSLA is better than PANN, the comparison between PSLA and PANN in Figure~\ref{fig-sota} shows that PANN performs better at higher coarse levels, which indicates PANN makes fewer false predictions on classes far from the target classes. The result in this section shows OmAP can provide more detailed evaluation results on different coarse-grained levels, and can better guide model comparison and performance analysis than mAP.

\begin{figure}[tbp] 
  \centering
  \vspace{-4mm}
  \includegraphics[page=1,width=1.0\linewidth]{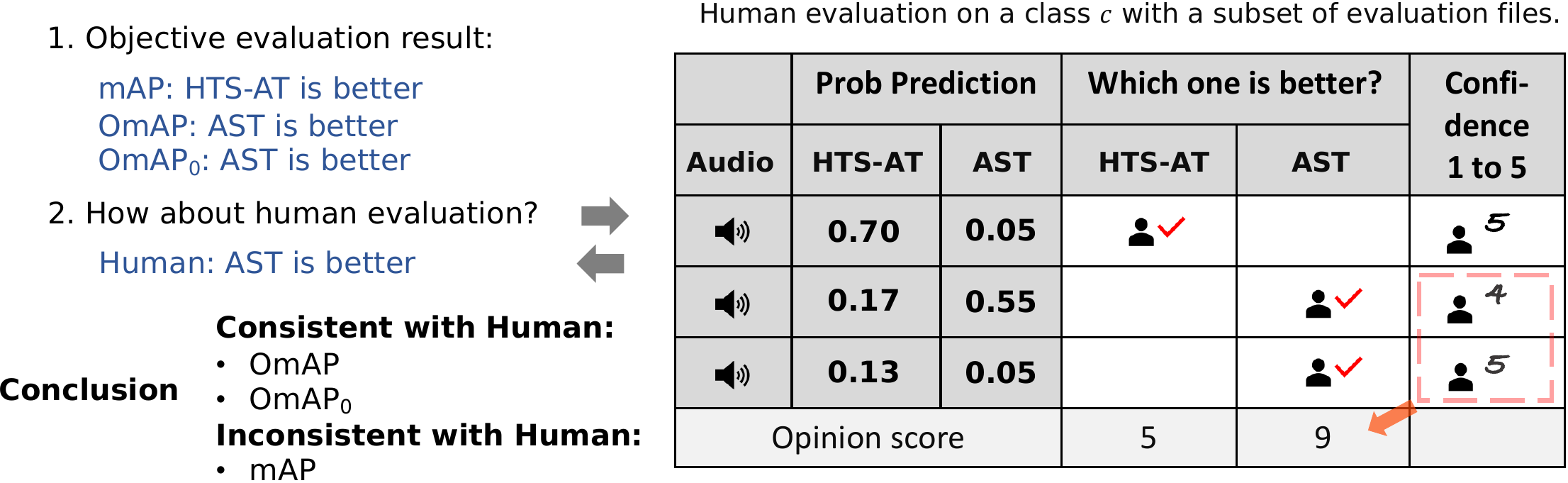}
  \caption{Comparing the objective evaluation result with the human evaluation score. The procedure in this figure will be performed multiple times with different class $c$ and subsets of evaluation files. Finally, for each objective metric, we calculate the statistic of agreement and disagreement with the human evaluation score to measure its consistency with human perception.}
  \vspace{-2mm}
  \label{fig-subjective_evaluation}

\end{figure}

\noindent
\textbf{Which metric is closer to human perception?}~~
By randomly sampling a class $c$ and a random subset of evaluation files to evaluate HTS-AT~\cite{chen2022hts} and AST~\cite{gong2021ast}, we observe 17\% of the results are inconsistent between mAP and OmAP on deciding which model is better. So, we design human evaluations on the inconsistent results as a reference to find out which metric is better.
As illustrated in Figure~\ref{fig-subjective_evaluation}, after listening to an audio, the participant need to choose which model makes better prediction and his/her confidence~(\num{1} to \num{5}). In our experiment, we found $94$\% of the answers are marked with the highest confidence.
We perform human annotation on $20$ different random classes and subsets of evaluation clips. 
For each class $c$ we randomly sample $30$ audio clips on the AudioSet evaluation subset both with and without the label of class $c$. We ensure there are at least $5$ audio clips with the label $c$. We also ensure for the clips without label $c$, at least half of them have AST probability estimation greater than $0.1$ on class $c$. We set these two constraints to ensure the majority of $30$ audio clips are relevant to class $c$, and have a reasonable proportion of positive and negative labels. We recruit four participants with audio processing backgrounds to perform this test. 
For each of the $20$ evaluation subsets, the participants are asked to determine which model is better, HTS-AT or AST, using the method shown in Figure~\ref{fig-subjective_evaluation}. On a subset of evaluation files, if an objective metric has the same result as the participant, we call this metric consistent with human perceptions.
The averaged consistency results on four participants are shown in Table~\ref{tab-subjective-evaluation}. Both OmAP and OmAP$_0$ show better consistency with human evaluation than mAP. 


\begin{table}[tbp]
\centering
\footnotesize
\begin{tabular}{cccc}
\toprule
    Evaluation metric   & mAP  & OmAP                 & OmAP${_0}$                              \\
\midrule
Consistency with human opinions  & 10.0\%   & \textbf{82.5\%}               & 62.5\%                       \\
\midrule
\multicolumn{1}{l}{} & \multicolumn{1}{l}{} & \multicolumn{1}{l}{} & \multicolumn{1}{l}{}
\end{tabular}
\vspace{-4mm}
\caption{The percent of agreement between human annotations and each objective evaluation metrics in \num{20} different trails.}
\label{tab-subjective-evaluation}
\vspace{-6mm}
\end{table}


\label{sec-exp-obceloss}

\begin{figure}[tbp] 
  \centering
  \vspace{-4mm}
  \includegraphics[page=1,width=0.8\linewidth]{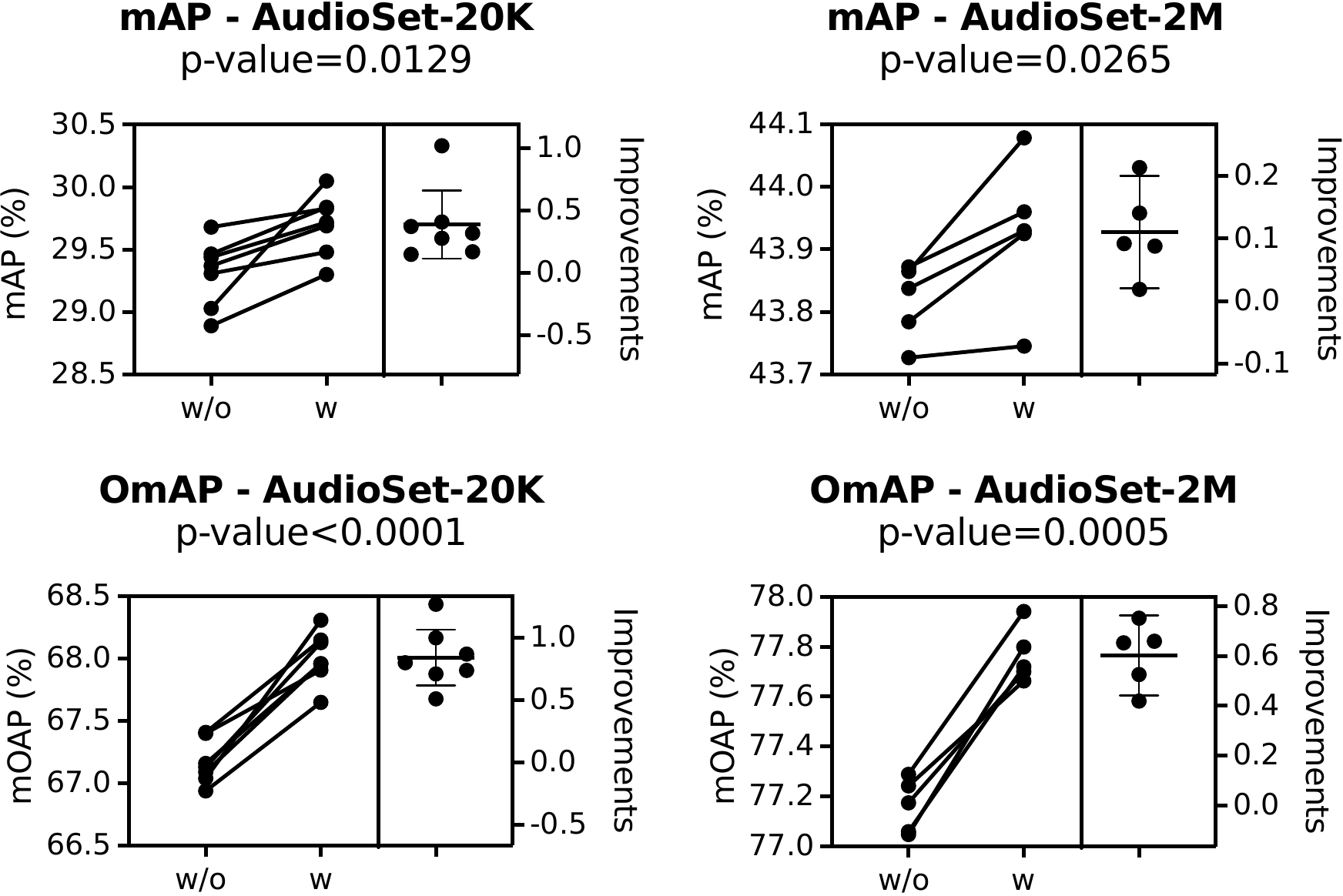}
  \caption{The mAP and OmAP without~(w/o) and with~(w) the OBCE loss. Both mAP and OmAP on AudioSet show improvement with more than $95\%$ confidence. The right half of each subfigure shows the values of improvements in the repeated experiments.}
  \label{fig-obce-compare}
  \vspace{-6mm}
\end{figure}

\noindent
\textbf{Improving audio tagging with OBCE loss}~~
We performed repeated experiments with different random seeds on both AudioSet-20K and AudioSet-2M with and without the OBCE loss. 
Our experimental results are shown in Figure~\ref{fig-obce-compare}, in which the experiments with the same seed are connected. We perform a paired t-test~\cite{dowdy2011statistics, tan2022naturalspeech} on the observed model performance with and without the OBCE loss.
The OmAP improvements with OBCE loss on the AudioSet-20K and AudioSet-2M are both statistically significant at more than $99\%$ confidence~($p\texttt{<}$\num{0.0001} and $p\vequalsignnospace$\num{0.0005}). This is expected because OmAP and OBCE are designed with similar motivations.
Surprisingly, we observe mAP can be improved on both datasets with $95\%$ confidence, which suggests reweighting the loss function based on ontology distance can also benefit model optimization and help the model make more accurate predictions. This might be because the reweighting in OBCE loss helps the model to learn the relation of classes, such as which classes are more (dis)similar to the target classes, and learn a better decision boundary.
We also conduct experiments on FSD50K and perform the same paired t-test. 
The result shows that the OBCE loss can improve the OmAP on FSD50K tagging task with \num{95}\% confidence~($p\vequalsignnospace0.03$), while our experiments do not show high confidence in improving mAP~($p\vequalsignnospace0.73$). 
This might be because the audio clips in FSD50K are more exhaustively labeled than AudioSet~\cite{fonseca2021fsd50k}, hence fewer labels are missing. Nevertheless, the improvements of mAP and OmAP with the OBCE loss presented in Figure~\ref{fig-obce-compare} indicate that ontology information is beneficial for audio tagging, which also suggests the proposed OmAP metric is preferable to the mAP.

As discussed in Section~\ref{sec-loss-function}, we introduce a distance power factor $\beta$ in the OBCE loss to explore the effect of non-linear distance between nodes. The parameter $\beta$ raises the elements of the distance matrix to a power and will affect model optimization. For example, a higher $\beta$ will make the difference between small and large values more prominent, which in turn makes the classes further from the target classes on the ontology have larger loss weight. When $\beta\vequalsign0$, the distance matrix becomes an all-one matrix, and the OBCE loss is reduced to the conventional BCE loss. The effect of $\beta$ on the AudioSet-20K is shown in Figure~\ref{fig-sota}. 
The OmAP improves roughly linearly with the increase of $\beta$, while mAP shows roughly a quadratic relation with $\beta$.
This indicates the OBCE loss might need a proper tuning of $\beta$ to achieve the best performance on OmAP and mAP.




\vspace{-3mm}
\section{Conclusions}
\label{Sec-conclusion}
In this paper, we proposed a new evaluation metric, ontology-aware mean average precision~(OmAP), which can evaluate model performance based on an intuitive class ontology. The multi-level coarse-grained evaluation scheme in OmAP provides more angles on model evaluation. 
Our human evaluation shows that OmAP is more consistent with human perceptions.
We also proposed a loss function, ontology-aware binary cross entropy~(OBCE) loss, that shows high confidence in improving both mAP and OmAP on AudioSet. 
The success of our proposed OBCE loss also supports our claim that OmAP is preferable to mAP as the audio tagging evaluation metric. Future work will be evaluating the OBCE loss on more SOTA models.

\vspace{-3mm}
\section{ACKNOWLEDGMENT}
\label{sec:ack}
This research was partly supported by the British Broadcasting Corporation Research and Development~(BBC R\&D), Engineering and Physical Sciences Research Council (EPSRC) Grant EP/T019751/1 ``AI for Sound'', and a PhD scholarship from the Centre for Vision, Speech and Signal Processing (CVSSP), Faculty of Engineering and Physical Science (FEPS), University of Surrey. For the purpose of open access, the authors have applied a Creative Commons Attribution (CC BY) license to any Author Accepted Manuscript version arising.

\bibliographystyle{IEEEtran}
\bibliography{strings,refs}

\end{document}